\documentclass[epj]{svjour}

\usepackage{graphicx}
\usepackage{fancyhdr}
\def\makeheadbox{{
 }}
\def\mytitle{My title} 
\def\myauthors{My name}
\def\mytype{My type of session}
\def\mysession{My session}
\def\mytitle{Searches in CDF Top quark samples} 
\def\myauthors{Susana Cabrera Urb\'{a}n}    
\def\mytype{Contributed Talk}    
\def\mysession{Alternatives}
\def\met{\mbox{${\hbox{$E$\kern-0.6em\lower-.1ex\hbox{/}}}_T$}} 
\begin{document}
\title{Looking for Signals of New Physics in the Top quark samples with the CDF detector.}


\author{Susana Cabrera \inst{1}
\thanks{\emph{Email:} Susana.Cabrera@ific.uv.es}
representing the CDF collaboration.} 
%
%
\institute{ \inst{1} IFIC, Institut de F\'{\i}sica Corpuscular, CSIC-Universitat de Val\`{e}ncia. 
Edifici Instituts d'Investigaci\'{o}, Apt. 22085, 46071, Val\`{e}ncia, Spain.}

%
\date{}

\abstract{
Twelve years after the discovery of the Top quark,
the CDF detector has collected large samples of $t\bar{t}$ events 
with integrated luminosities up to 1-1.7 $fb^{-1}$, 
where we are able to probe our knowledge of the Standard Model Top quark.
This article will foccuss on those analysis results dedicated to 
find new physics in the Top quark production: measurement of
relative fractions of Top pair production mechanisms, searches 
for a massive resonant state decaying to $t\bar{t}$ pairs as well as
a heavy Top-like quark decaying as the Standard Model Top quark.
Besides its production, the Top quark decay is also an interesting probe
to find Physics Beyond the Standard Model by measuring 
possibly anomalous values of the helicity fractions of the 
W boson in the decay $t{\rightarrow}W({\rightarrow}{\ell}{\nu}_{\ell})b$,
by searching for flavour changing neutral current decays $t{\rightarrow}Zq$
or by testing the Top quark charge to be $+2/3$ or $-4/3$.
}
\PACS{
      {PACS-key}{discribing text of that key}   \and
      {PACS-key}{discribing text of that key}
     } 

%
\maketitle
\section{Introduction}
\label{sec:intro}

The Top quark is the most massive fundamental particle observed by experiment~\cite{cdfdiscovery,d0discovery}.
The last measured value, $m_{top}=170.9{\pm}1.8~GeV/c^2$~\cite{mtop}, very close to the electroweak scale, suggests 
that the coupling to a Higgs boson would be close to unity and leads to speculate about a special role that the 
Top quark could play in the electroweak symmetry breaking.
Having a lifetime $\hbar/\Gamma_t \sim 5\times 10^{-25}$~s, an order of magnitude shorter than 
the typical strong interaction time-scale for binding of quarks into hadrons, $\hbar/\Lambda_{\mathrm QCD} \sim 3 \times 10^{-24}$~s,
the Top quark decays before hadronization, and the spin information is directly transferred to the decay products allowing us
to test the Standard Model (SM) theory with the $t\bar{t}$ final state kinematics. 

The Top quark signal has been reestablished in the enlarged data sample ($\sim$ 1-1.7 $fb^{-1}$) 
collected by the CDF detector through the measurement of  ${\sigma}(p\bar{p}{\rightarrow}t\bar{t})$ in all $t\bar{t}$ 
final states using independent top quark samples with different sensitivities to non-SM effects. 
The golden $t\bar{t}$ sample corresponds to the {\it lepton plus jets} final state. With one W boson 
decaying hadronically and the other one leptonically, this $t\bar{t}$ signature has a medium yield of events and
a moderate level of background. Is in the {\it lepton plus jets} channel where CDF obtains 
the best measured value: ${\sigma}_{\rm CDF}^{t\bar{t}}=8.5{\pm}1.1$ pb. This value is combined with
those from other $t\bar{t}$ final states obtaining a combined measured value 
of ${\sigma}_{\rm CDF}^{t\bar{t}}=7.3{\pm}0.9$ pb, which  is in good agreement with the
QCD NLO predictions, ${\sigma}_{\rm NLO}^{t\bar{t}}=6.7{\pm}0.8$ pb for $m_{top}=175~GeV/c^2$. 
Both theoretical and experimental results reach an accuracy of $\sim$ 12 $\%$. 
The $t\bar{t}$ {\it lepton plus jets} selection criteria demand the presence of only one electron (or muon) 
with $E_{T}(P_{T})>20~GeV/c$, four o more jets with $E_{T}>20~GeV$, missing transverse energy to 
account for the non-interacting neutrino greater than 30 $GeV$ and at least one {\it b-tagged}
jet with secondary vertex. In an integrated luminosity of 1.12 $fb^{-1}$, 231 candidates were observed
with a background of 21 $\pm$ 6 and a $t\bar{t}$ signal expectation of 207 $\pm$ 17 assumming
 ${\sigma}^{t\bar{t}}=8.2$ pb. Most of the analyses presented in this article utilize this sample.


\section{New physics in the production of $t\bar{t}$ pairs.}

\subsection{ ${\sigma}(gg{\rightarrow}t\bar{t})/{\sigma}(p\bar{p}{\rightarrow}t\bar{t})$ measurements.}
At $\sqrt{s}=1.96$ TeV the SM predicts that the fraction of $t\bar{t}$ pairs produced via gluon fussion $t\bar{t}^{gg}$ 
(quark annihilation $t\bar{t}^{q\bar{q}}$ ) is 15 (85) $\pm$ 5 $\%$, which is affected by high uncertainties.
The measurement of ${\sigma}(gg{\rightarrow}t\bar{t})/{\sigma}(p\bar{p}{\rightarrow}t\bar{t})$
is sentitive simultaneously to both new Top production mechanisms hidden by new Top decay modes \cite{GluonFussion}.

\begin{figure*}
\begin{center}
\begin{tabular}{cccc}
\includegraphics[width=0.25\textwidth,height=0.28\textwidth,angle=0]{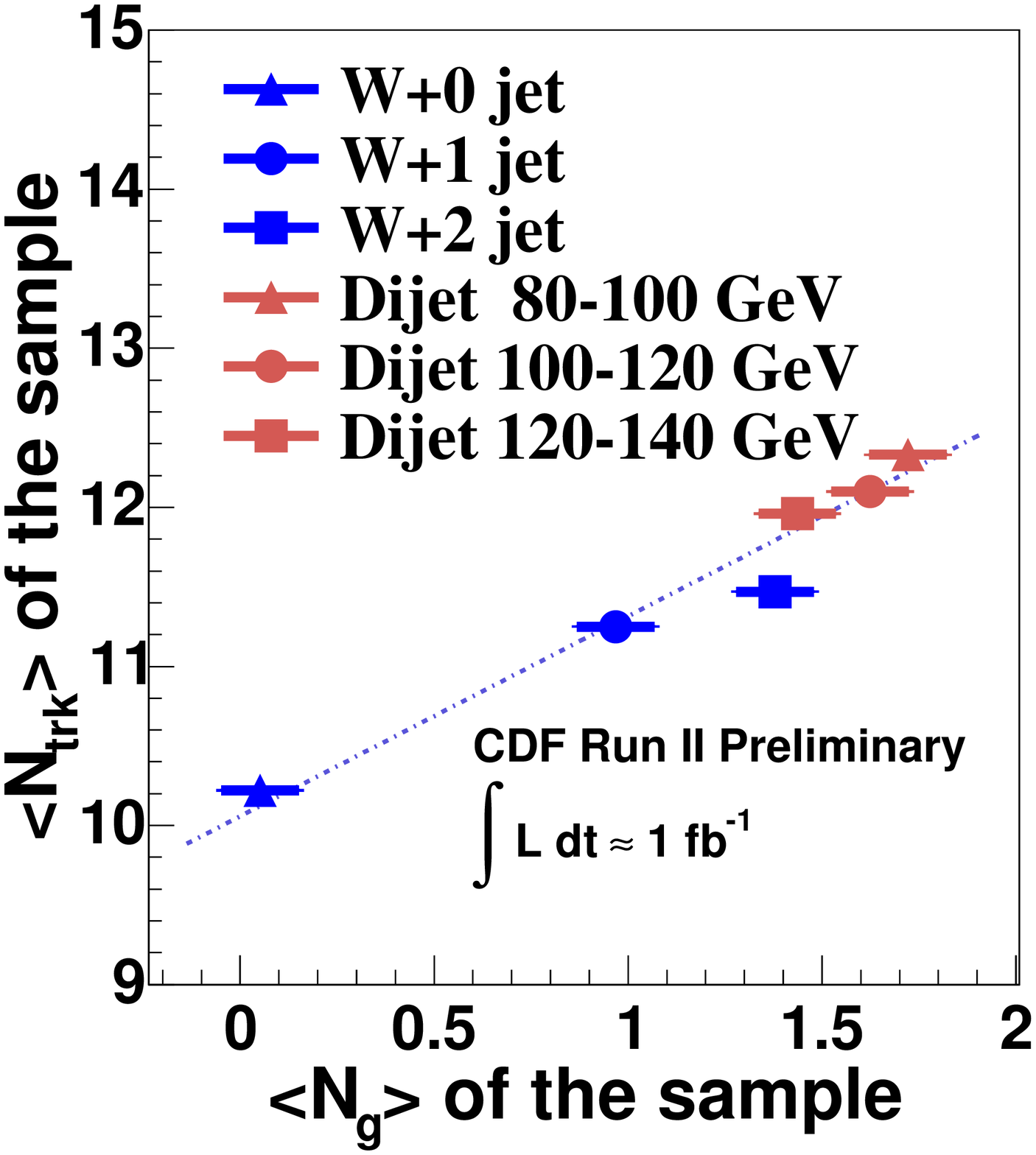}
\includegraphics[width=0.25\textwidth,height=0.28\textwidth,angle=0]{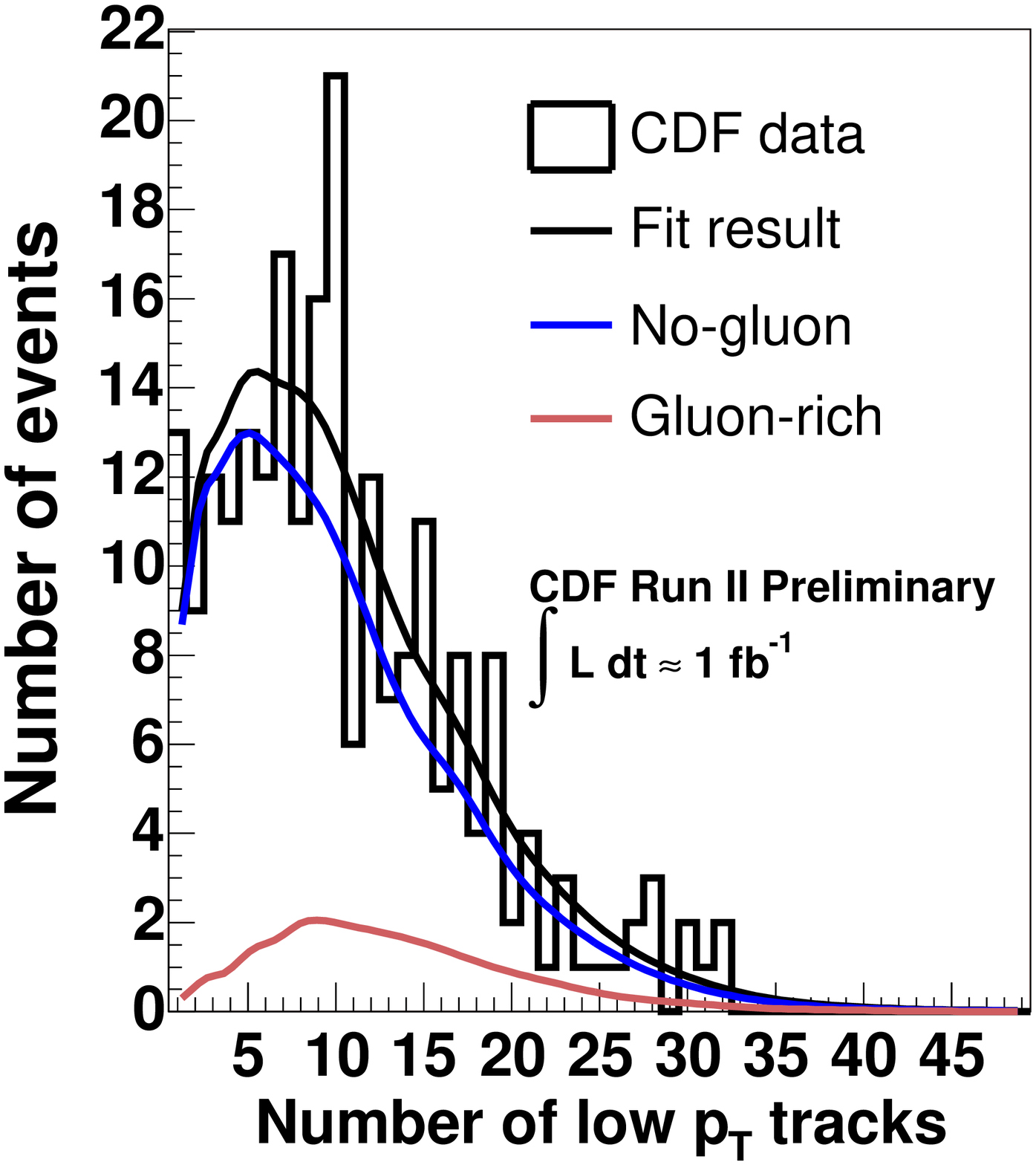}
\includegraphics[width=0.25\textwidth,height=0.28\textwidth,angle=0]{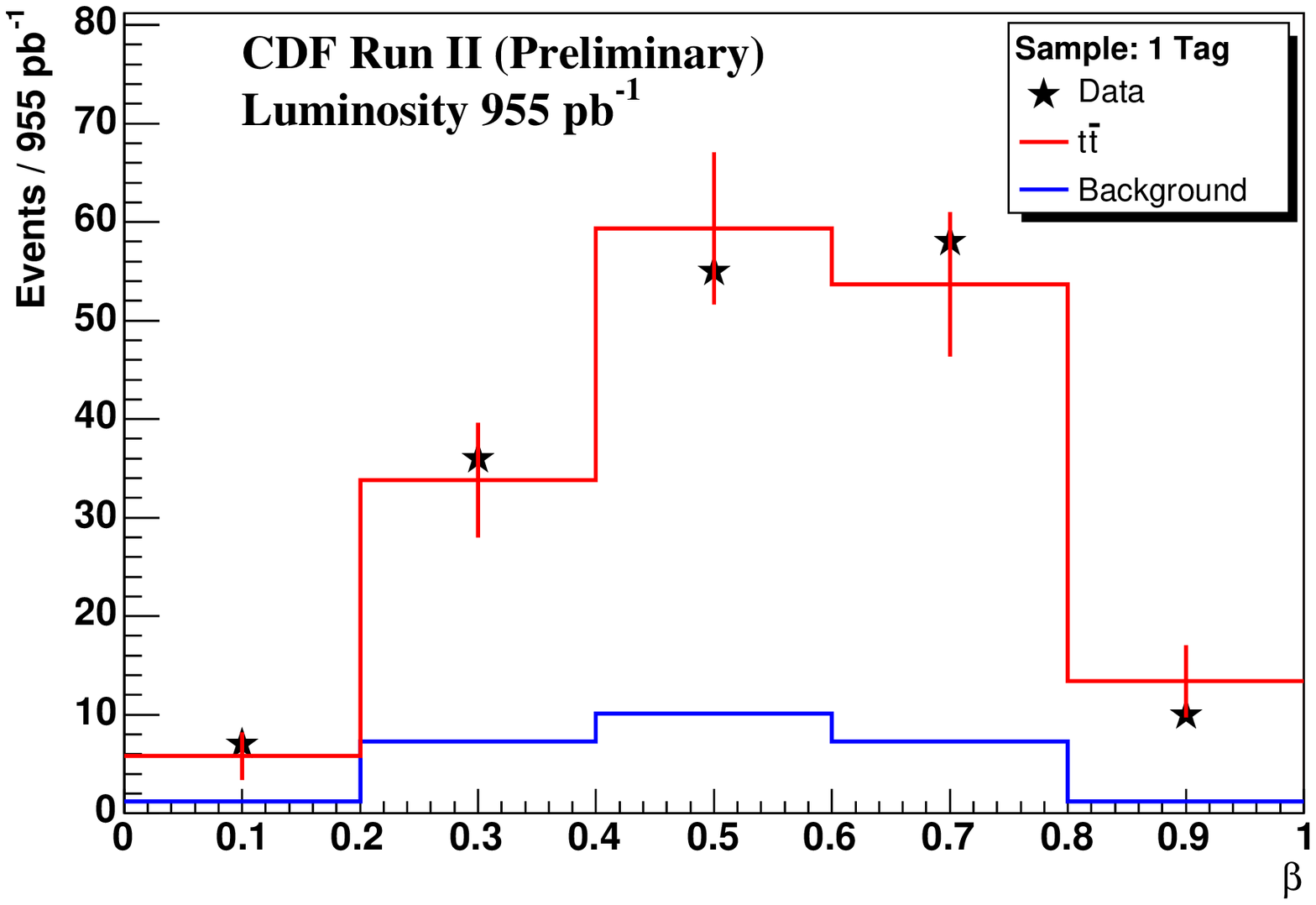}
\includegraphics[width=0.25\textwidth,height=0.28\textwidth,angle=0]{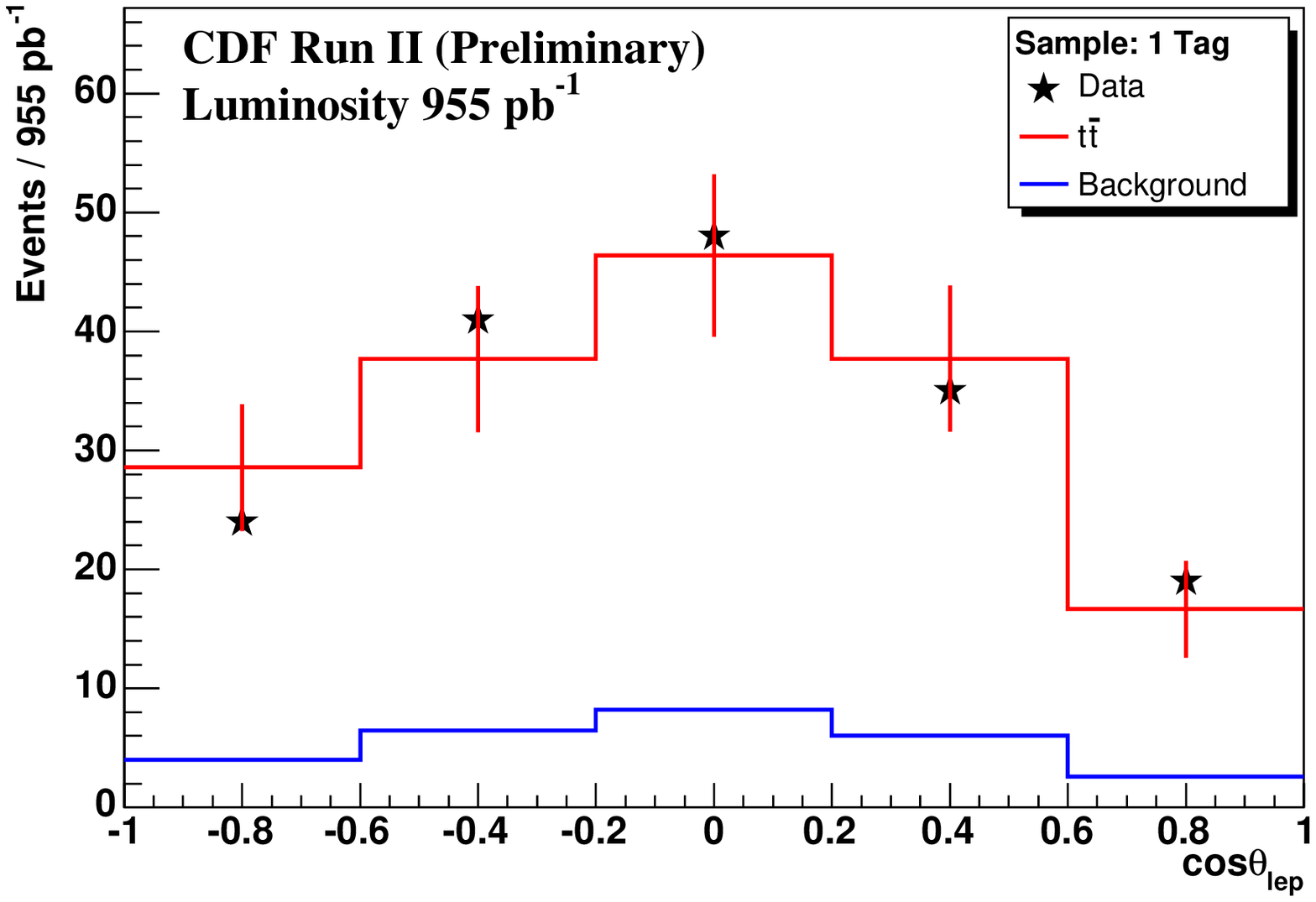}
\end{tabular}
\end{center}
\caption{
(a) $<N_{trk}>$ versus $<N_{g}>$ correlation. (b) low $P_{T}$ tracks templates, non-gluon (DIJET 80-100 GeV) and gluon-rich (W+0 jets),
data and fit result. (c) Speed of the Top quark in the $t\bar{t}$ reference frame. (d) $cos{\theta}_{lepton}$. }
\label{fig:ggttbar}       
\end{figure*}

A first experimental approach \cite{cdf8724} exploits the fact that $gg{\rightarrow}t\bar{t}$ events tend to have more underlaying event activity with respect to 
$q\bar{q}{\rightarrow}t\bar{t}$ events and therefore more soft charged particles. The correlation between the average number of low $P_{T}$ tracks 
$<N_{trk}>$ versus the mean number of gluons $<N_{g}>$ is calibrated using W+n-jets and dijet data and MC processes.
$<N_{trk}>$ is measured from data by counting tracks with $P_{T}=$ 0.3-2.9 $GeV/c$ outside the cones of the existing jets in the event. 
$<N_{g}>$ is derived from MC by counting gluons among the two immediate incoming and outgoing partons in the generator.
The calibrated correlation allows to defined low $P_{T}$ tracks templates
both non-gluon (DIJET 80-100 GeV) and gluon-rich (W+0 jets) (see first two plots in figure \ref{fig:ggttbar}). A sample of W+n-jets
{\it b-tagged} data is fitted to these two templates leading to a measured value of 
${\sigma}(gg{\rightarrow}t\bar{t})/{\sigma}(p\bar{p}{\rightarrow}t\bar{t})
=0.07{\pm}0.14(stat)$ ${\pm}0.07(syst)$.

An alternative method \cite{cdf8811} discriminates between $t\bar{t}^{gg}$ and $t\bar{t}^{q\bar{q}}$ using a Neural Network feeded
variables that provide kinematic information from the production in the $t\bar{t}$ rest frame: (a) the speed of the Top quark
(see 3rd plot in figure \ref{fig:ggttbar}) and (b) the angle between the Top quark and the incoming partons. 
The spin correlation information from the decay is also considered in the Network using the angles between the decay 
products and the {\it ``off-diagonal''} basis, direction in the $t\bar{t}$ rest frame where the like spin components vanish on average
(see 4th plot in figure \ref{fig:ggttbar} with the angle of the lepton and the {\it ``off-diagonal''} basis). 
The data template obtained with the output of the Network is fitted to MC $t\bar{t}^{q\bar{q}}$ and $t\bar{t}^{gg}$ templates to extract an upper
limit in the true fraction of $t\bar{t}^{gg}$ events at the 95 $\%$ confidence level: 
${\sigma}(gg{\rightarrow}t\bar{t})/{\sigma}(p\bar{p}{\rightarrow}t\bar{t})<0.61$.

\subsection{New production mechanisms.}
\label{sec:zprime}
\subsubsection{ Search for resonant $t\bar{t}$ production. }
\vspace{-0.2cm}
CDF has search for a $Z^{'}$ resonance with narrow width, 1.2 $\%$
of its mass, with the same couplings as the Z boson and with no
resonant interference with the s-channel gluon production \cite{cdf8675}.
Such search is sensitive to new resonant producion mechanisms 
predicted by different exotic models: Extended Gauge theories \cite{egg}, 
KK states of the gluon and the Z \cite{kkgZ} and Topcolor \cite{topcolor}.
The reconstructed mass of the $t\bar{t}$ event, $M_{t\bar{t}}$,
is performed with the same kinematic fitter used to measured the 
Top mass \cite{mtopprd}. The data is fitted with a binned likelihood 
method to three $M_{t\bar{t}}$ templates: SM $t\bar{t}$, 
non $t\bar{t}$ SM background and a resonance $Z^{'}{\rightarrow}t\bar{t}$ 
in a range of masses: 450$<M_{Z^{'}}<$ 900 $GeV/c^2$.
For $M_{Z^{'}}>700$ $GeV/c^2$, an upper limit to the production
of a $Z^{'}$ resonance is set: ${\sigma}(Z^{'}{\rightarrow}t\bar{t})<0.7$
$pb^{-1}$ at the 95 $\%$ confidence level (see second plot in figure \ref{fig:newphysprod}).

\subsubsection{ Search for a heavy top $t^{'}{\rightarrow}Wb$. }
\begin{figure*} 
\begin{center}
\begin{tabular}{cccc}
\includegraphics[width=0.25\textwidth,height=0.28\textwidth,angle=0]{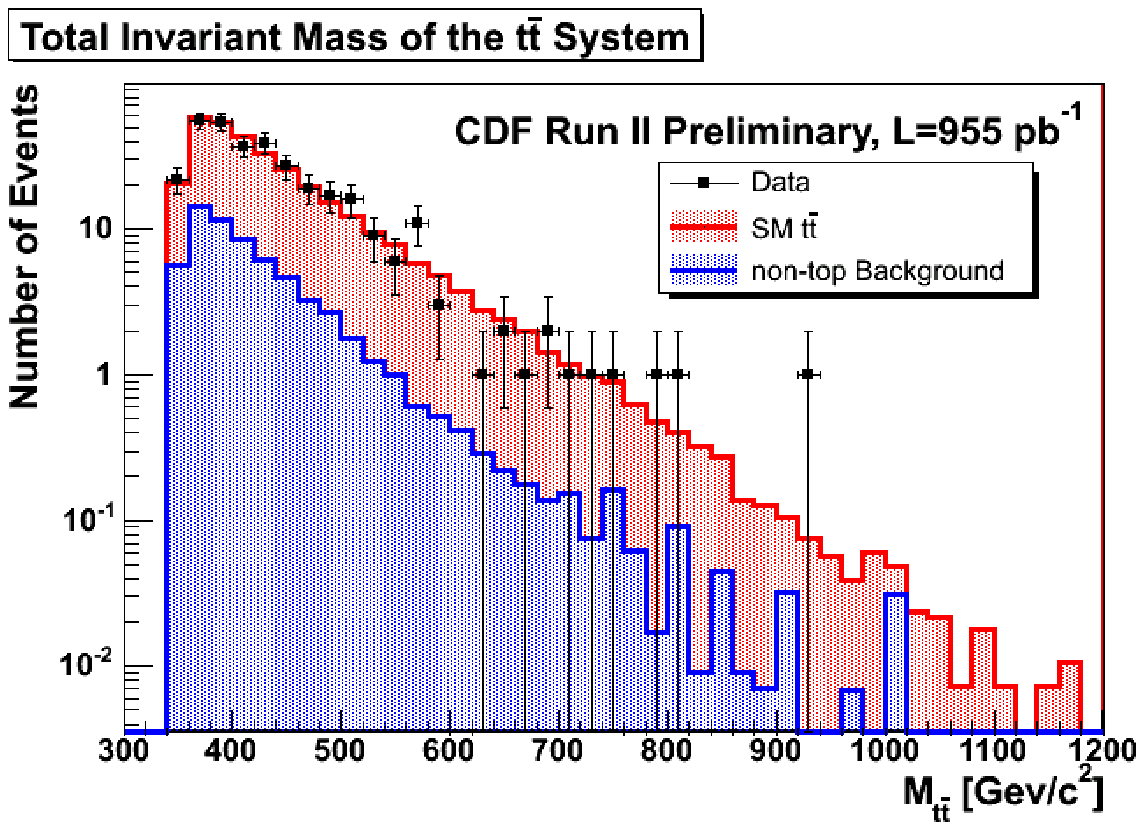}
\includegraphics[width=0.25\textwidth,height=0.28\textwidth,angle=0]{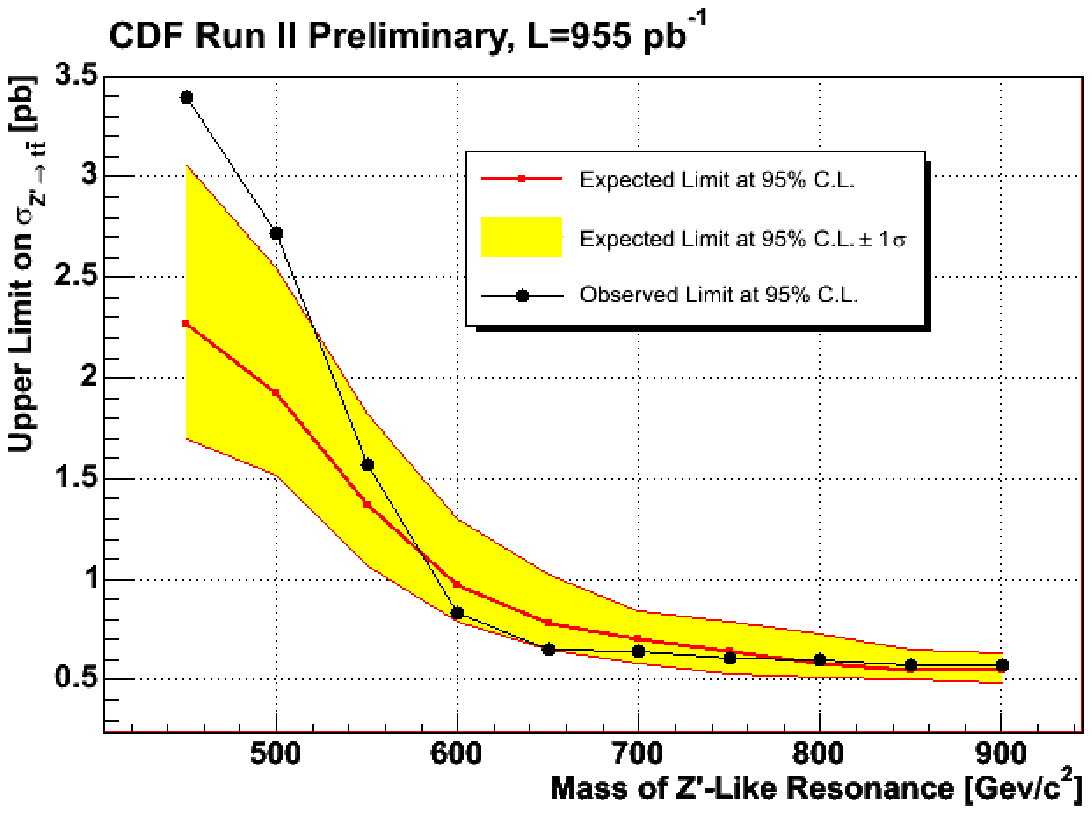}
\includegraphics[width=0.25\textwidth,height=0.28\textwidth,angle=0]{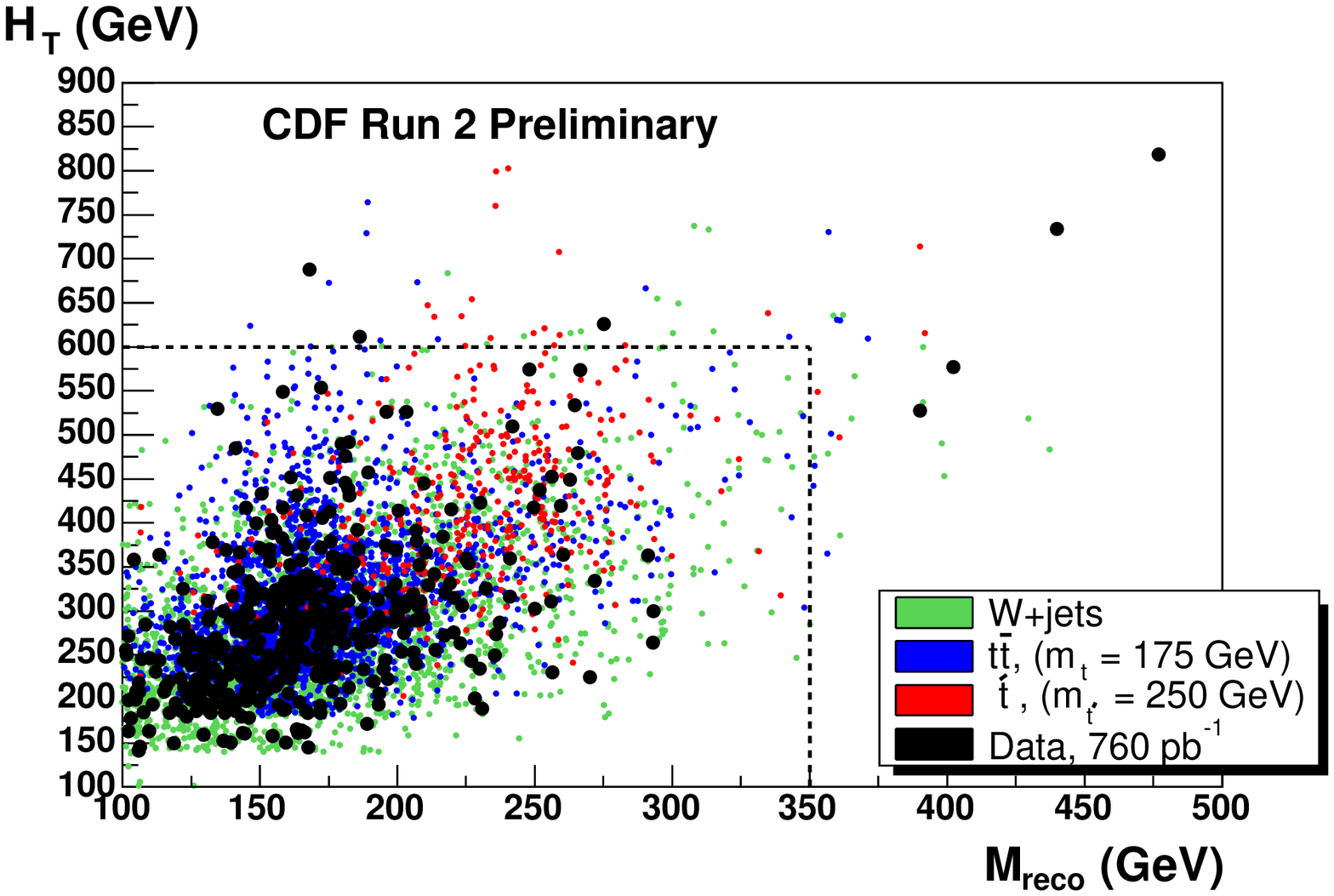}
\includegraphics[width=0.25\textwidth,height=0.28\textwidth,angle=0]{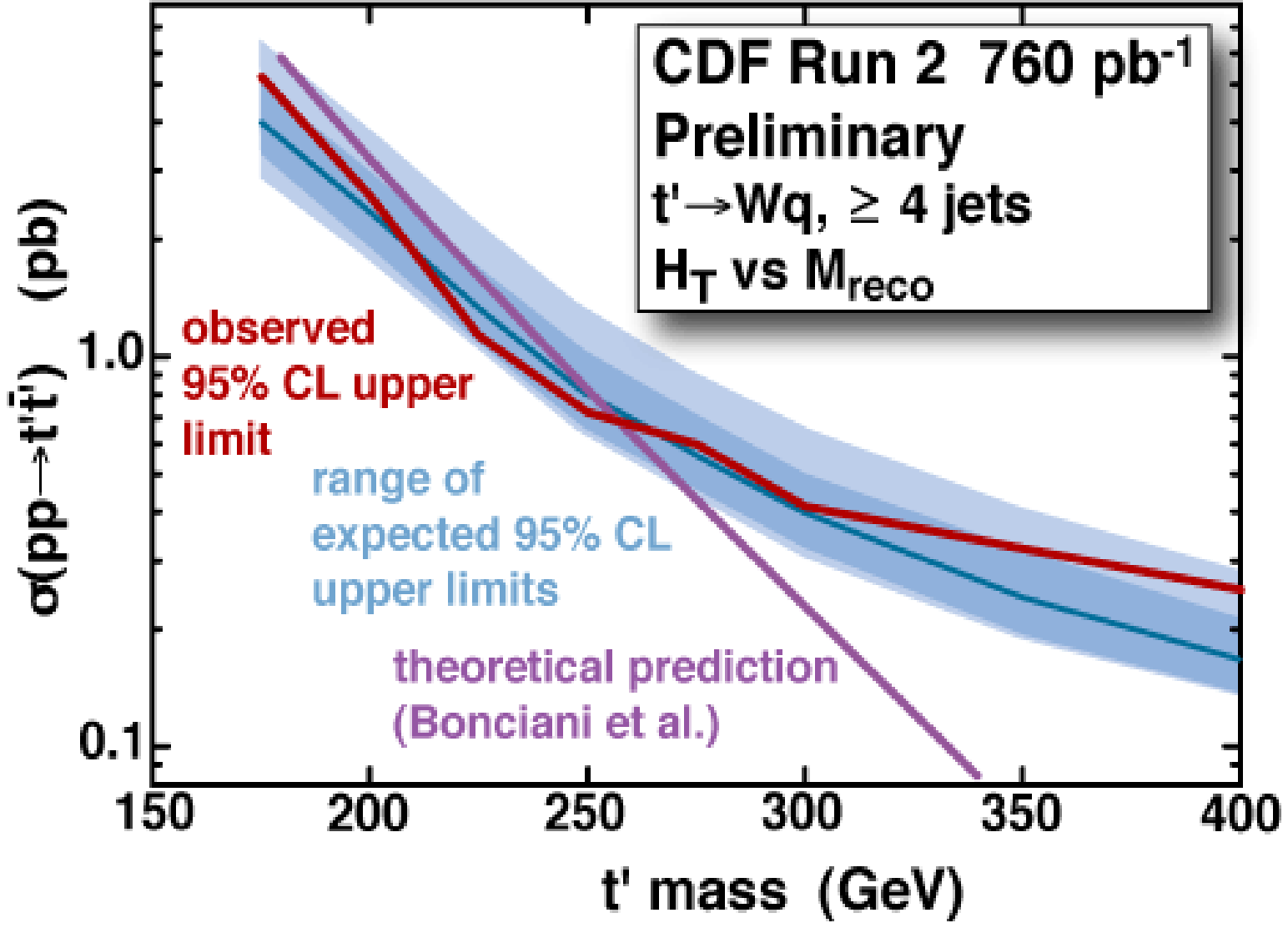}
\end{tabular}
\end{center}
\caption{ (a) Reconstructed mass of the $t\bar{t}$ system. (b) Upper limit on ${\sigma}_{Z^{'}{\rightarrow}{t\bar{t}}}$ at 95$\%$ C.L. versus the $Z^{'}$ mass. (c) $H_{T}$ versus reconstructed  $t\bar{t}$ mass. (d) Upper limit on ${\sigma}_{t^{'}\bar{t}^{'}}$ at 95$\%$ C.L. versus
the $t^{'}$ mass.}
  
\label{fig:newphysprod}       
\end{figure*}

\vspace{-0.2cm}
Some theoretical scenarios point out that a heavier fourth generation with $m_{Z}/2<m_{f}<m_{Higgs}$ would be consistent
with the existing precision electroweak data \cite{hepph0102144},\cite{hepph0111028}.
In particular, Two-Higgs-doblet models and N=2 SUSY models could accomodate a heavier 4th 
fermion generation with a $t^{'}$ quark heavier than the Top quark, and also 
produced strongly in pairs and with a decay mode $t^{'}{\rightarrow}Wb$ \cite{hepph0102144}. 
Other scenarios lead to the hypothesis of the existance of such $t^{'}$ extra quark, for instance
``beautiful mirrows'' models \cite{hepph0109097} and non-minimal little Higgs models \cite{than}.
The experimental approach follow by CDF to search for this $t^{'}$ quark \cite{cdf8495}
has consisted of fitting the data to three MC templates: SM $t\bar{t}$, non-$t\bar{t}$ background
and $t^{'}\bar{t}^{'}$ with a 2 dimensional binned likelihood in two variables: $M_{t\bar{t}}$,
built as in section \ref{sec:zprime} and the total transverse energy in the event $H_{T}= \sum_{jets} E_{T}+E_{T,{\ell}}+\met$. The Standard Model fourth-generation $t^{'}$ quark with a mass below 256 $GeV/c^2$ is excluded at the 95 $\%$ confidence level ( see 4th plot in figure \ref{fig:newphysprod}).

\section{New physics in the Top quark decay.}
\subsection{ Search for anomalous measured Top quark properties.}
\subsubsection{ W boson helicity fractions in the decay chain $t{\rightarrow}W({\rightarrow}{\ell}{\nu}_{\ell})b$. }

In general, the t-W-b coupling can have form-factors of type V-A, V+A and 
magnetic moment ~\cite{kane}. Only the V-A form-factors are allowed in the SM.
The presence of non-SM couplings in the Wtb vertex could significantly 
modify the polarization of the top quark decay products indicating the presence 
of new phenomena, for instance left-right symmetric models or models with mirrow fermions ~\cite{theory:vpa1,theory:vpa2,theory:vpa3}.
In the SM, the spin-$\frac{1}{2}$\ top quark decays via
the charged current weak interaction to a spin-$1$ $W^{+}$ boson
and a spin-$\frac{1}{2}$\ $b$ quark with a branching fraction above 99$\%$.
In the limit $m_b{\rightarrow}0$, the $b$ quark has left-handed ($-1/2$) polarization (helicity) and
the $W^{+}$ boson can only have either longitudinal (zero) or left-handed ($-1$) polarization.
The right-handed ($+1$) polarization is forbidden.
The fraction $f^0$ of $W^+$ bosons with longitudinal polarization is predicted
at leading order in perturbation theory to be $f^0 = m^2_t/(2m^2_W+m^2_t)=0.70$.

In CDF, two experimental approaches (\cite{cdf8449} and \cite{cdf8938}) infer the W boson helicity fractions from the measurement of 
the observable cos(${\theta}^*$), where ${\theta}^*$ is the angle between the charged lepton in the W boson rest frame and the W boson 
in the top rest frame: 
$\frac{1}{{\Gamma}} \frac{d\Gamma}{d\cos\theta^*}
\hspace*{-0.35mm}=\hspace*{-0.15mm}f^0\frac{3}{4}\hspace*{-0.15mm}(1\hspace*{-0.95mm}-\hspace*{-0.65mm}\cos^2\hspace*{-0.35mm}\theta^*)\hspace*{-0.35mm}
\hspace*{-0.35mm}+\hspace*{-0.15mm}f^+\frac{3}{8}(1\hspace*{-0.45mm}-\hspace*{-0.65mm}\cos\hspace*{-0.35mm}\theta^*)^{2}\hspace*{-0.35mm}
\hspace*{-0.35mm}+\hspace*{-0.15mm}f^-\frac{3}{8}(1\hspace*{-0.45mm}+\hspace*{-0.65mm}\cos\hspace*{-0.35mm}\theta^*)^{2}\hspace*{-0.35mm}$

\begin{figure*}
\begin{center}
\begin{tabular}{cccc}
\includegraphics[width=0.30\textwidth,height=0.28\textwidth,angle=0]{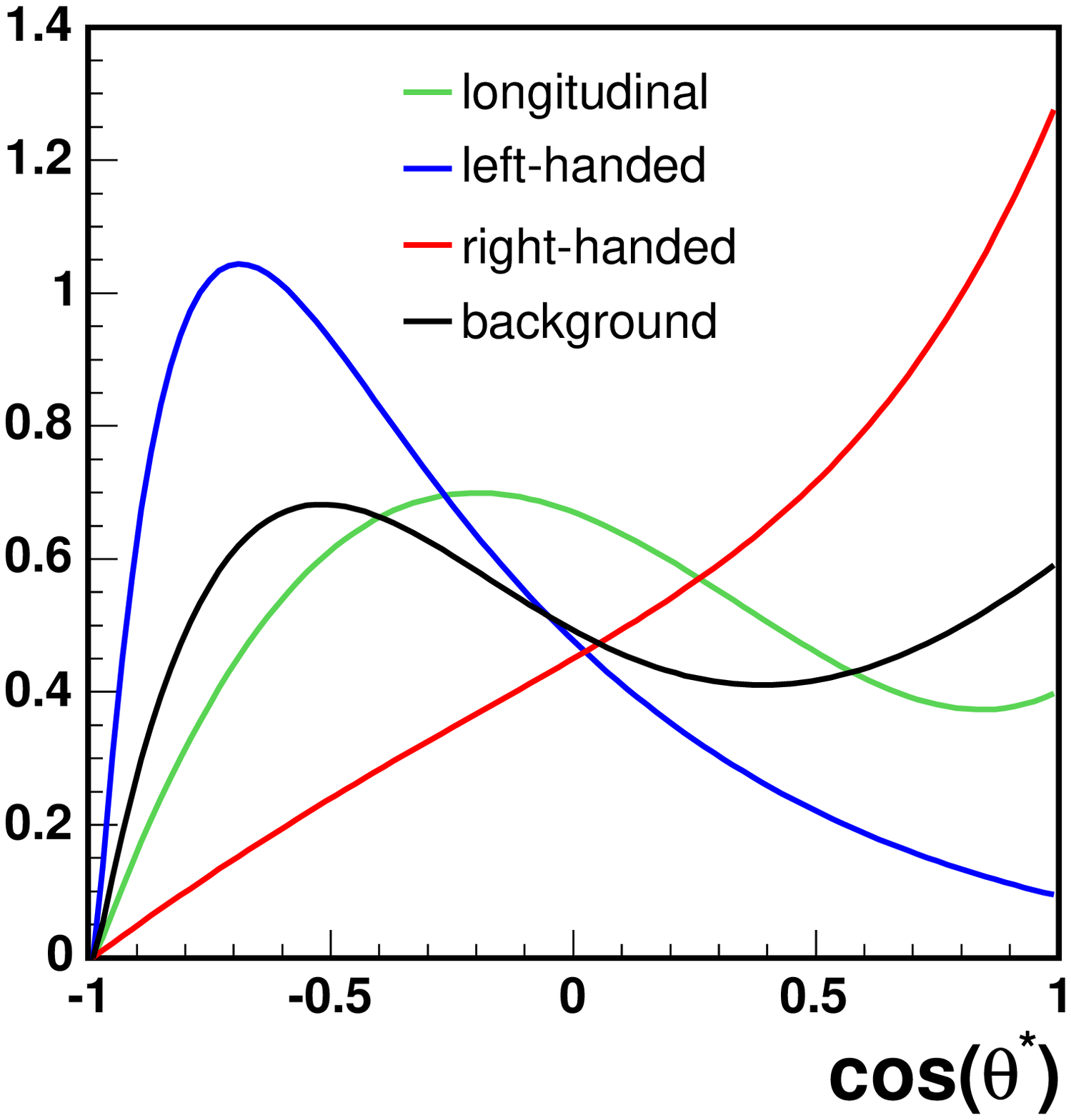}
\includegraphics[width=0.25\textwidth,height=0.25\textwidth,angle=0]{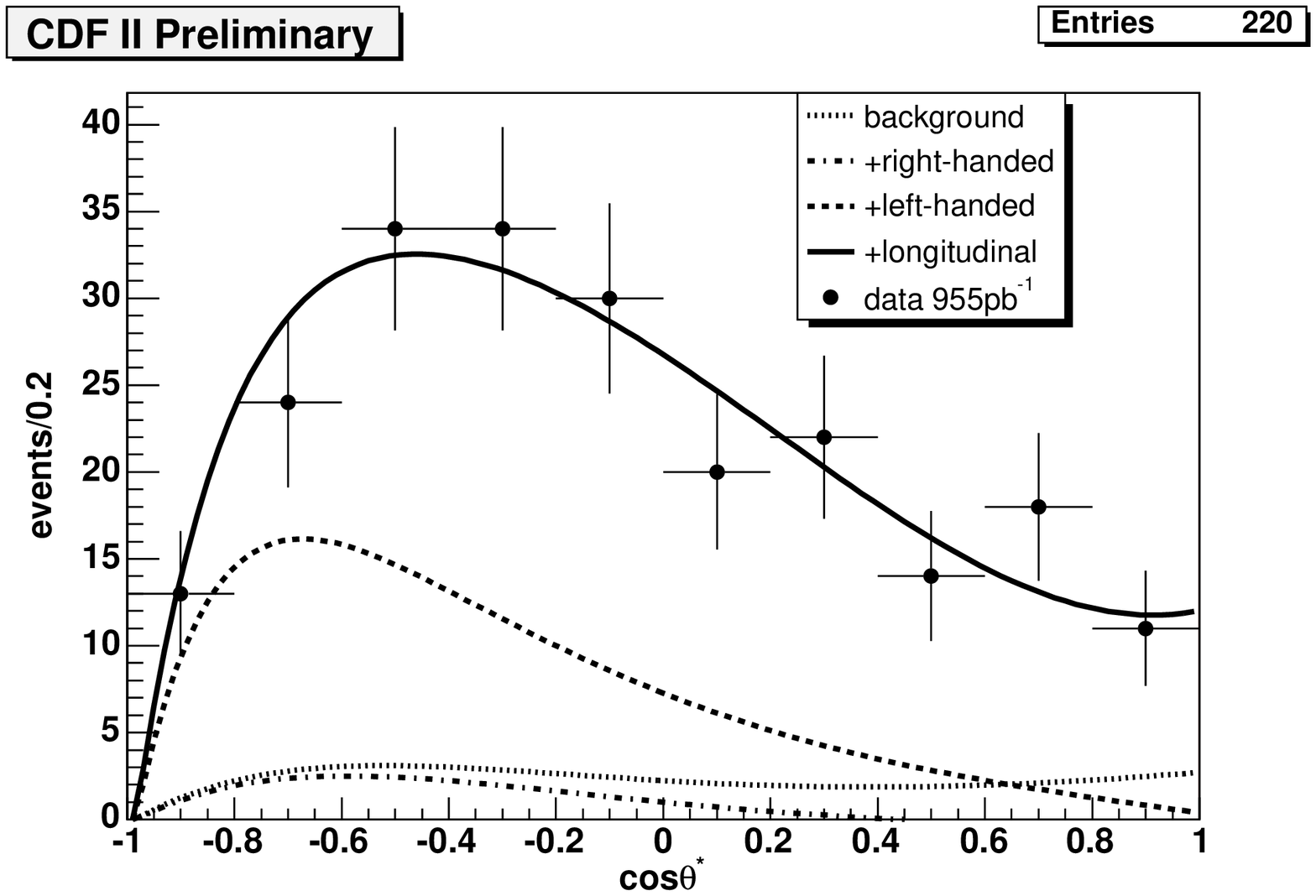}
\includegraphics[width=0.25\textwidth,height=0.25\textwidth,angle=0]{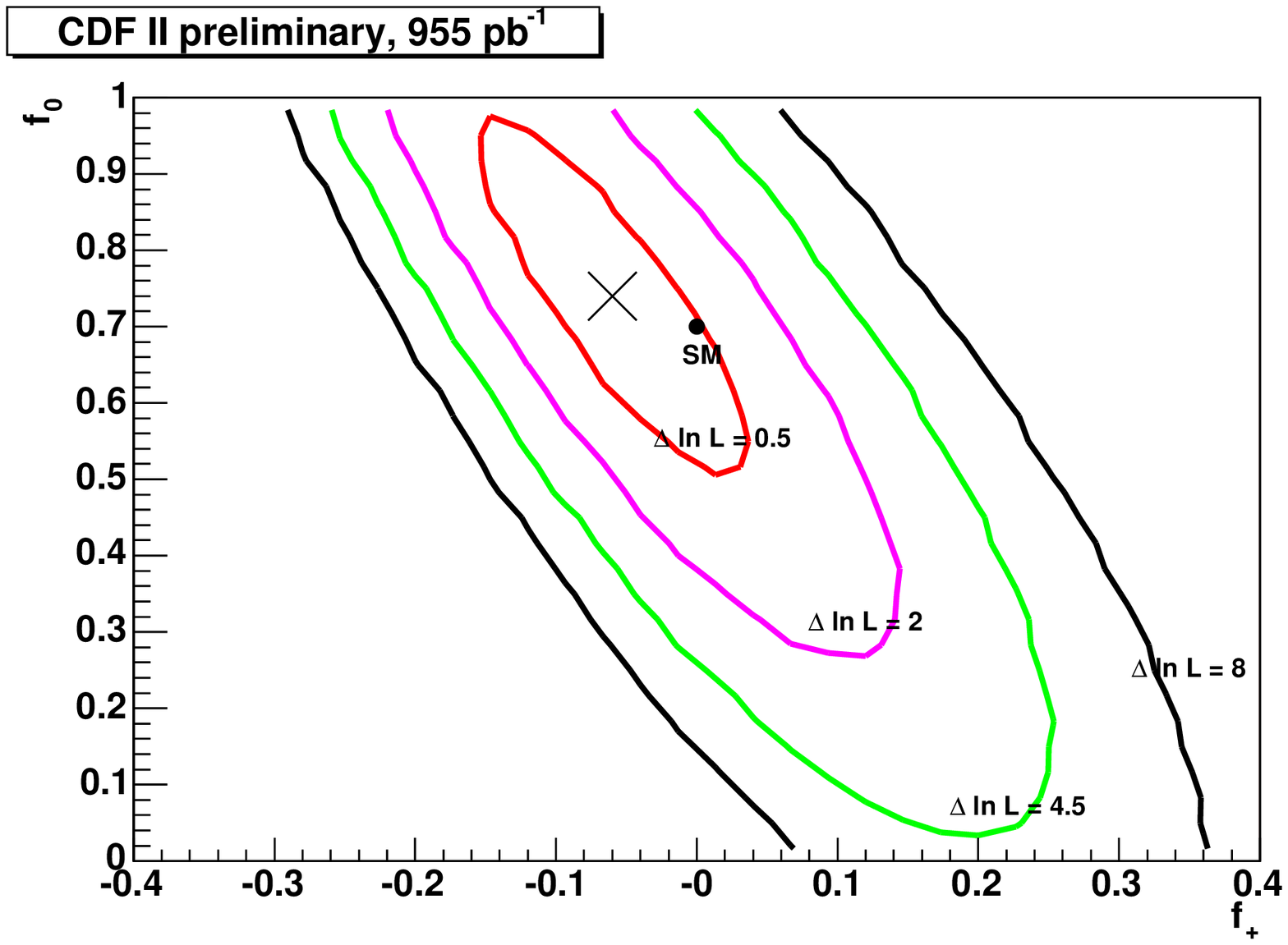}
\includegraphics[width=0.25\textwidth,height=0.25\textwidth,angle=0]{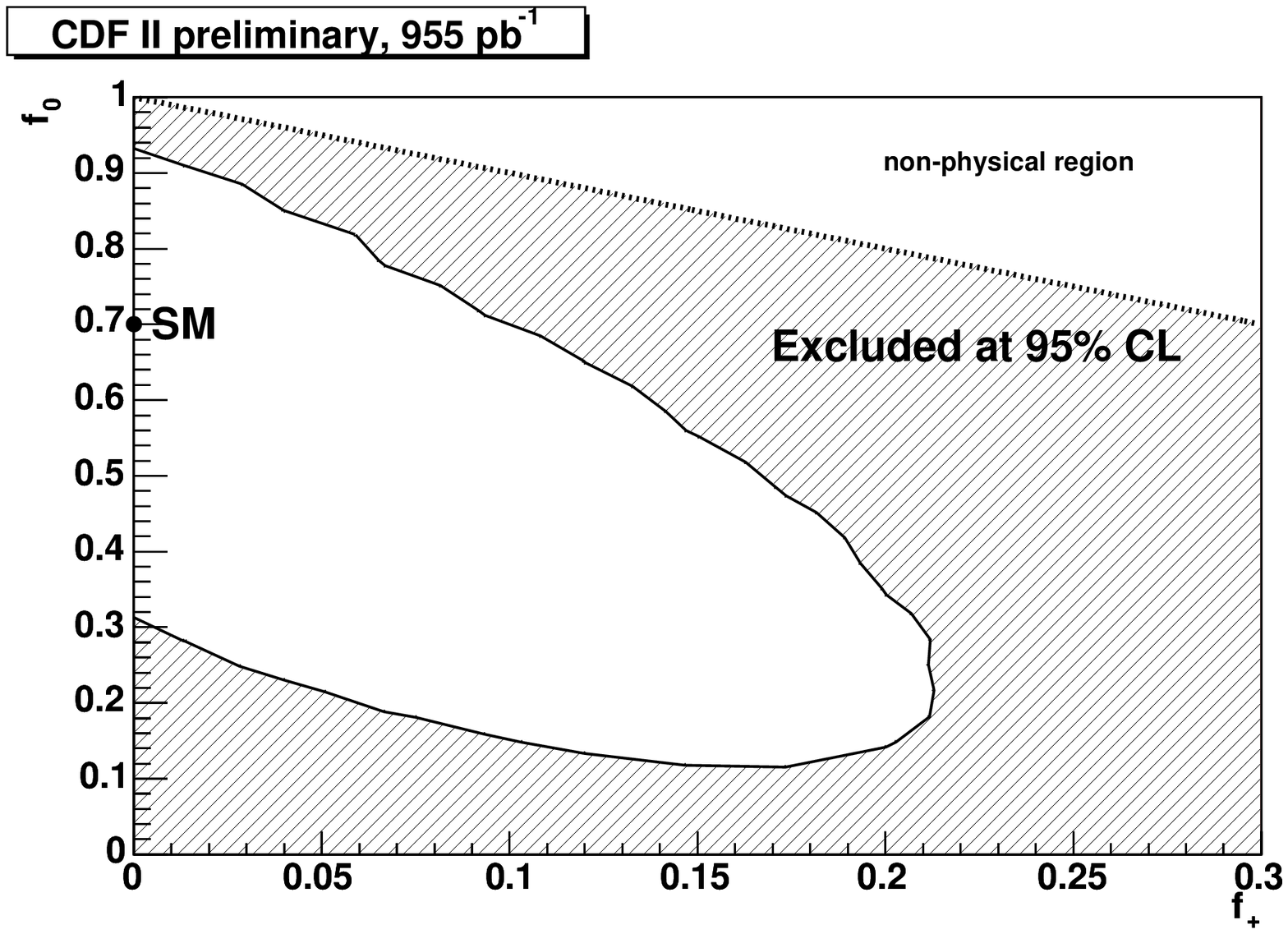}
\end{tabular}
\end{center}
\caption{(a) $t\bar{t}$ righ-(left-)handed , longitudinal and background parameterized templates.(b) Observed $cos(\theta^*)$ in data overlaid with background, $t\bar{t}$ right-handed, left-handed and longitudinal accumulatively added as determined by the simultaneous fit in $f^0$ and $f^+$. (c) Contour plots of constant likelihood in $f^0$ vs. $f^+$ plane.(d) Area $f^0$ vs. $f^+$ excluded by the simultaneous fit.}
\label{fig:whel1}       
\end{figure*}

\begin{figure*}
\begin{center}
\begin{tabular}{cccc}
\includegraphics[width=0.25\textwidth,height=0.28\textwidth,angle=0]{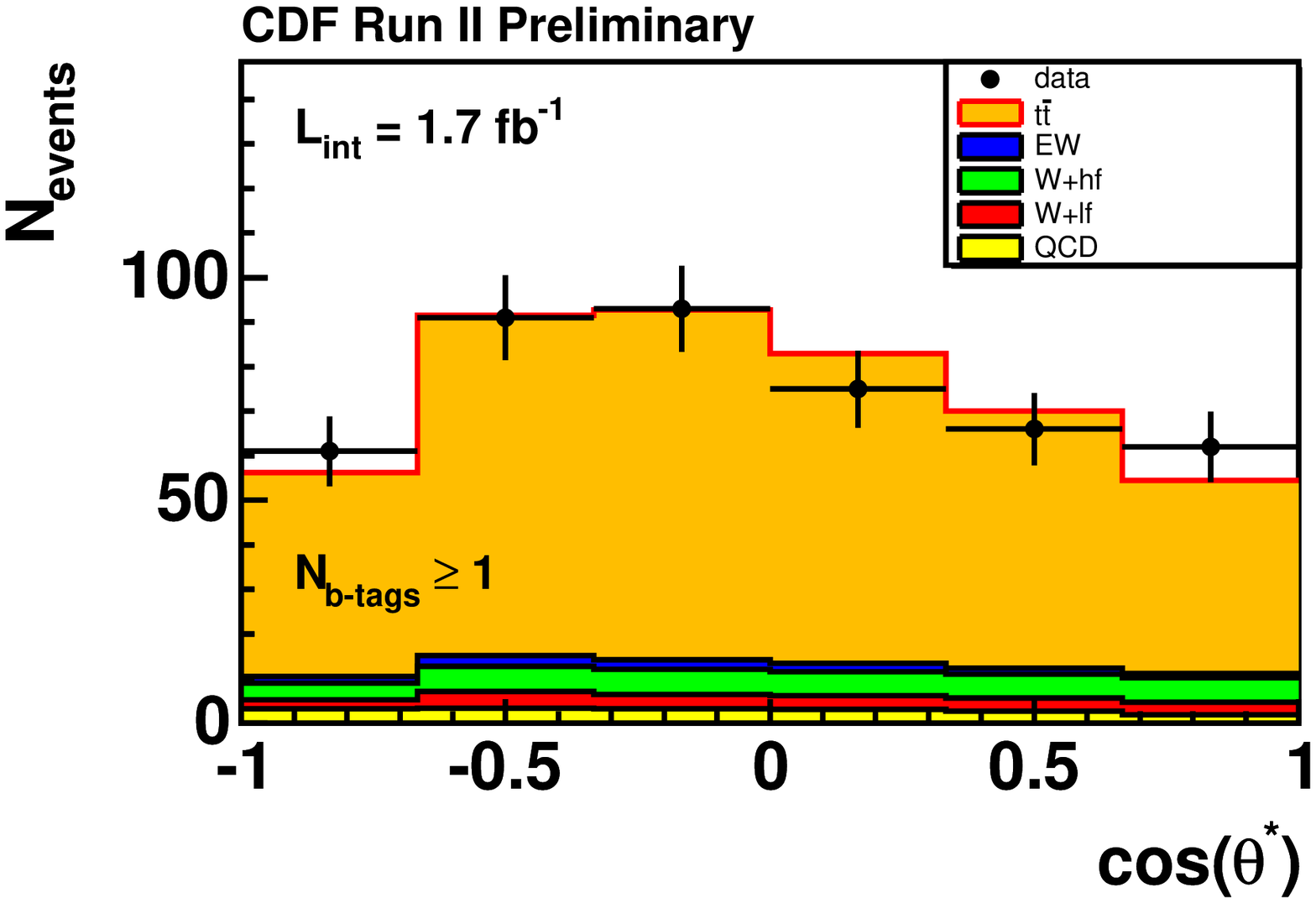}
\includegraphics[width=0.25\textwidth,height=0.28\textwidth,angle=0]{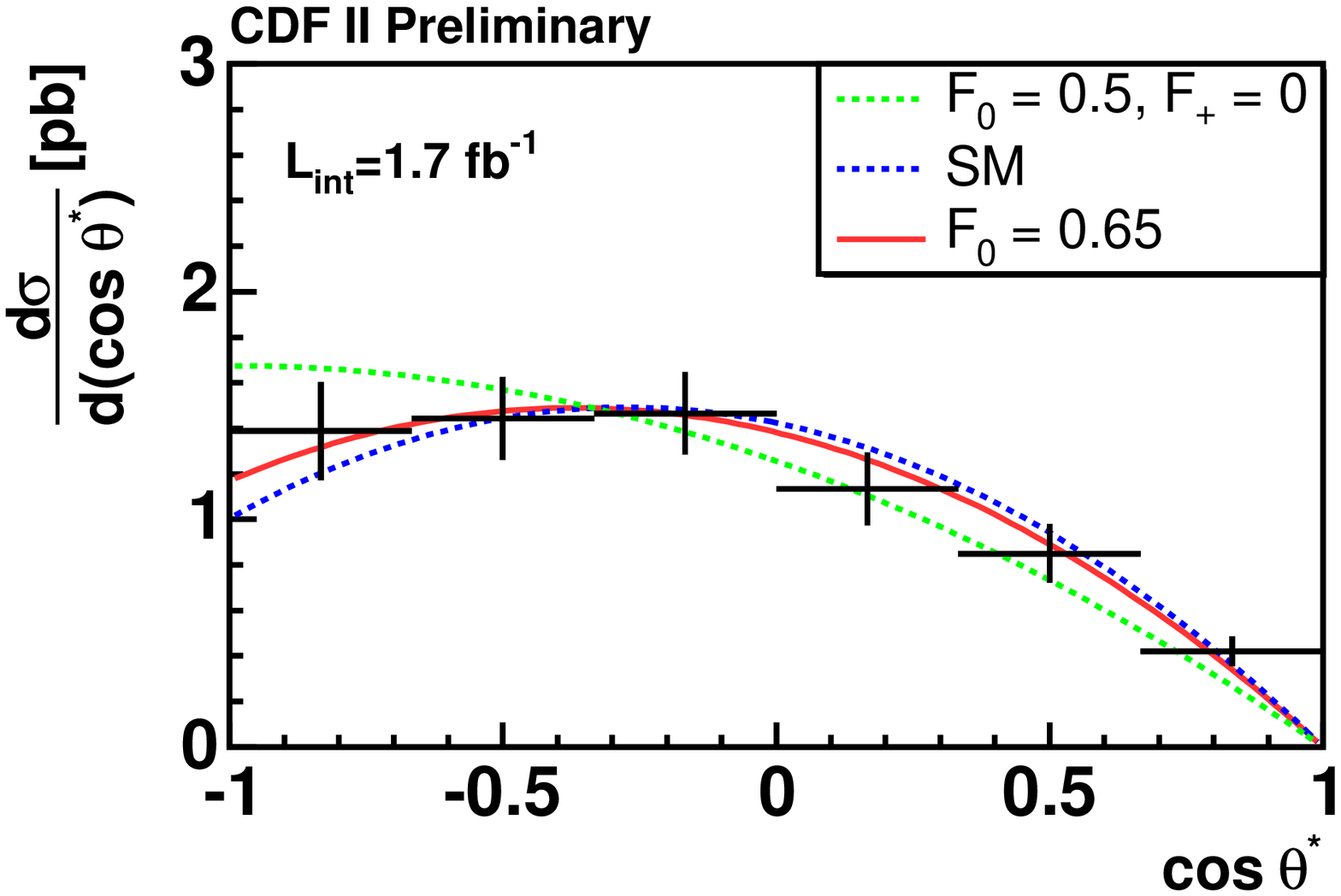}
\includegraphics[width=0.25\textwidth,height=0.28\textwidth,angle=0]{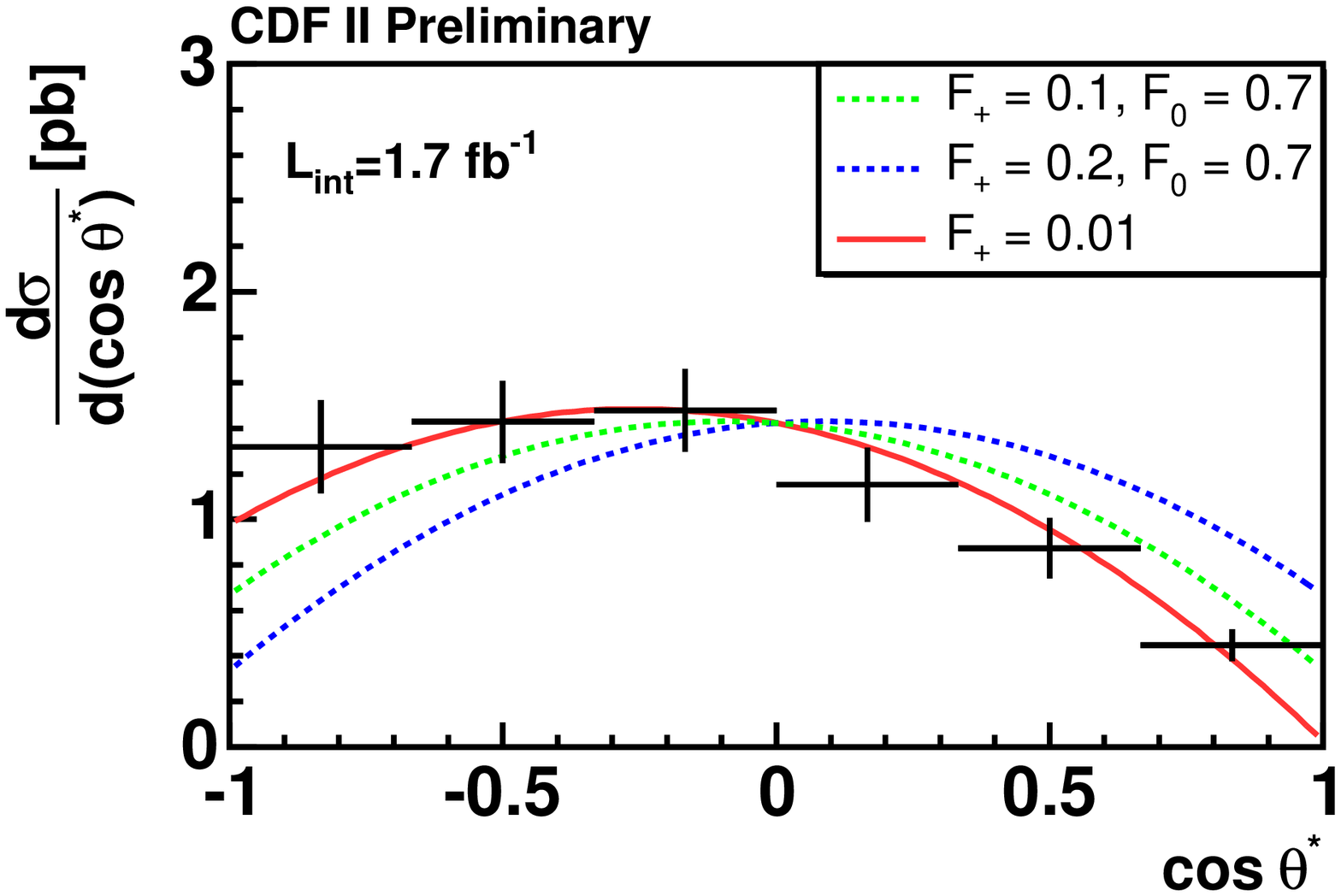}
\includegraphics[width=0.25\textwidth,height=0.28\textwidth,angle=0]{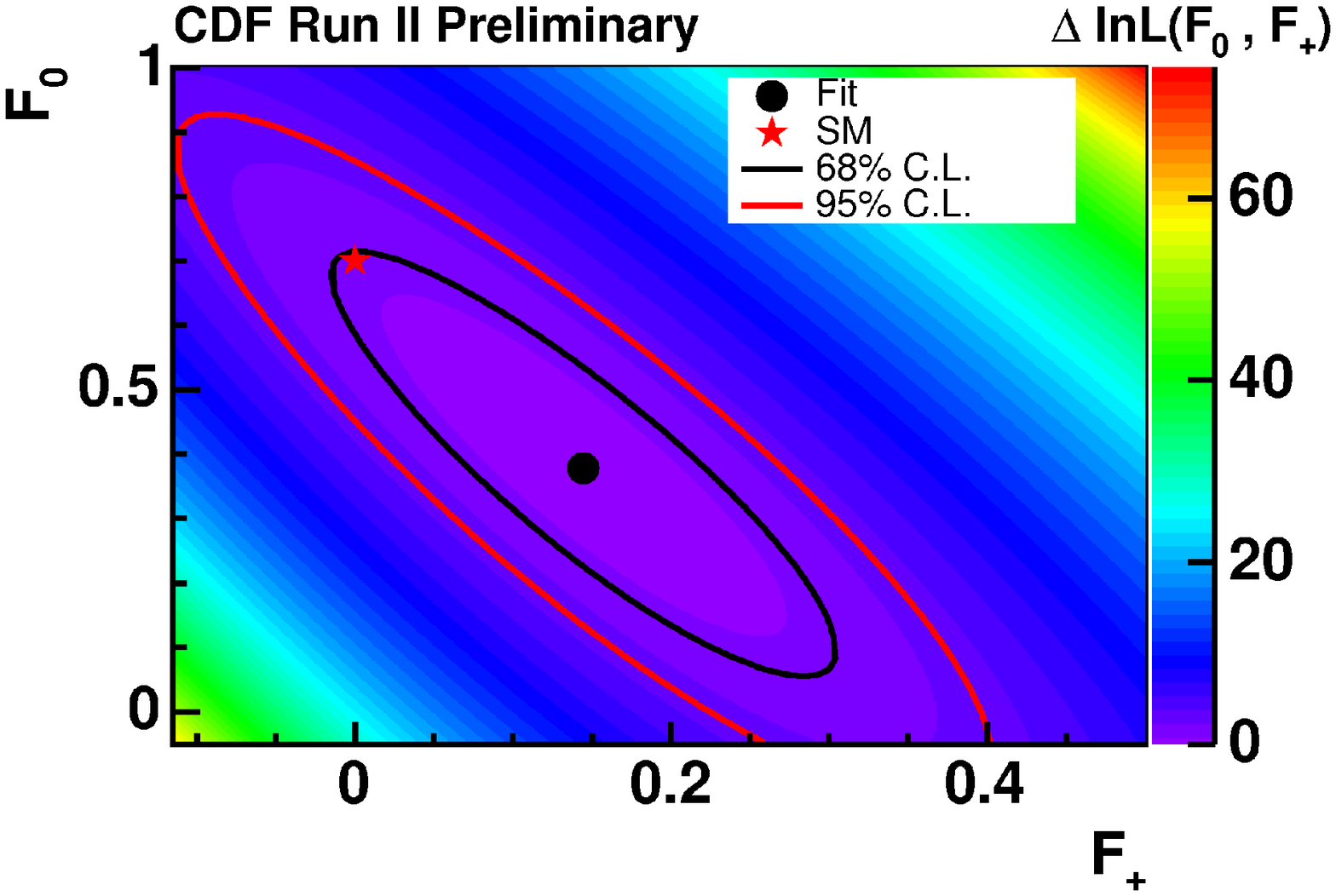}
\end{tabular}
\end{center}
\caption{(a) Reconstructed $cos(\theta^*)$ observable measured with data overlaid with MC $t\bar{t}$ and background. (b) and (c) 
$cos(\theta^*)$ corrected for detector efficiency and resolution effects and compared with theoretical curves for different set of $f^0$ and $f^+$ values. 
(d) Contour plots of constant likelihood in $f^0$ vs. $f^+$ plane.}
\label{fig:kal}       
\end{figure*}

The analysis \cite{cdf8449} (\cite{cdf8938}) uses $\sim$ 1 (1.7)  $fb^{-1}$ $t\bar{t}$ {\it lepton plus jets} sample.
Both techniques start with the fully reconstruction of the $t\bar{t}$ kinematics with a $\chi^2$ minimization
as described in \cite{mtopprd}. Then, cos(${\theta}^*$) is extracted by boosting the lepton and the top
quark to the W rest frame and finally $f^0$ and $f^+$ are extracted. In \cite{cdf8449} the data is fitted
to longitudial, right-handed, left-handed $t\bar{t}$ and background templates using an unbinned likelihood fitter.
The templates are parameterized with third order polinomial time exponential functions (see figure \ref{fig:whel1}). 
In \cite{cdf8938} the data is fitted to signal and background templates with a binned likelihood fitter. The templates
are derived from theoretical principles and detector efficiency and resolution effects are included (see figure \ref{fig:kal}). 
In Table \ref{table:whel} the results from \cite{cdf8449} and \cite{cdf8938} are compared.

\begin{table*}
\centering
\begin{tabular}{lllll}
\hline\noalign{\smallskip}
Ref & $L~fb^{-1}$ & $f^0$ & $f^+$ & $f^+<$ 95$\%$ C.L.\\
\noalign{\smallskip}\hline\noalign{\smallskip}

\cite{cdf8449} & 1. &  0.7 &  0.06 $\pm$ 0.06 (stat) $\pm$ 0.03 (syst) &  0.11 \\
\cite{cdf8938} & 1.7 &  0.7 &  0.01 $\pm$ 0.05 (stat) $\pm$ 0.03 (syst) &  0.12 \\

\cite{cdf8449} & 1. &  0.61 $\pm$ 0.12 (stat) $\pm$ 0.06 (syst) & 0. &  - \\ 
\cite{cdf8938} & 1.7 &  0.65 $\pm$ 0.10 (stat) $\pm$ 0.06 (syst) & 0. &  - \\

\cite{cdf8449} & 1. &  0.74 $\pm$ 0.25 (stat) $\pm$  0.06 (syst) & -0.06 $\pm$ 0.10 (stat) $\pm$  0.03 (syst) & ( 4th plot fig. \ref{fig:whel1} ) \\
\cite{cdf8938} & 1.7 &  0.38 $\pm$ 0.22 (stat) $\pm$  0.07 (syst) &  0.15 $\pm$ 0.10 (stat) $\pm$  0.04 (syst) & ( 4th plot fig. \ref{fig:whel2}) \\

\noalign{\smallskip}\hline
\end{tabular}
\caption{Results of analyses \cite{cdf8449} and \cite{cdf8938} for three fit cases: 1-D fit in $f^+$ fixing $f^0=0$ in first and second raws,
2-D fit in $f^0$ fixing $f^+=0$  in third and fourth raws and the simultaneous fit in $f^0$ and $f^+$ in fifth and sixth raws.}
\label{table:whel}       
\end{table*}

\subsubsection{Top quark charge.}

\vspace{-0.2cm}

The experimental hypothesis of a Top quark charge $\frac{-4e}{3}$
could lead to the existance of a new exotic quark part of a heavier four generation \cite{topcharge}.
The key ingredient of the CDF analysis \cite{cdf8782} 
is the algorithm used to get the flavour of the b-jet, that uses the sum of the charge
of the tracks within the jet cone averaged with the proyection of the track momentum
along the jet axis. 
The result is consistent with a charge $\frac{2e}{3}$ and the exotic quark 
hypothesis is excluded with 81 $\%$ confidence.
 
\vspace{-0.2cm}
\subsection{ New decay modes: FCNC $t{\rightarrow}Zq$}
\vspace{-0.2cm}
In the SM, the Top quark FCNC decays are highly suppressed by GIM mechanism
and CKM suppression. Beyond the SM scenarious enhance Top quark FCNC decays providing observable
branching fractions for instance $t{\rightarrow}Zq$ \cite{aguilar}.
CDF has recently preformed a blind search for the process $t\bar{t}{\rightarrow}Z({\rightarrow}ee,{\mu}{\mu})qW({\rightarrow}q\bar{q}^{'})b$.
The leptonic decay of the Z boson provide a clean signature and the hadronic decay of the W boson a larger branching fraction of events.
In order to optimize the analysis, the signal has been separatde in two regions: with zero and $\ge1$ b-tags. 
The events selection has been optimized for the best expected limit by cutting in the variable mass $\chi^2=(\frac{m_{W_{rec}}-m_{W_{PDG}}}{{\sigma}_{W_{rec}}})^2
+ (\frac{m_{t{\rightarrow}Wb_{REC}} - m_{t_{PDG}}}{{\sigma}_{t{\rightarrow}Wb}})^2 -(\frac{m_{t{\rightarrow}Zq_{REC}} - m_{t_{PDG}}}{{\sigma}_{t{\rightarrow}Zq}})^2  $
After optimization the unblinding leads to an observed number of events consistent with background.
An upper limit of $BR({\rightarrow}Zq)<10.6\%$ was set at 95 $\%$ confidence level, improving the previous limit:
$BR({\rightarrow}Zq)<13.7\%$ from non-observation of $e^+e^-{\rightarrow}tZ$ at LEP,L3.

\vspace{-0.4cm}
\section{Conclusions}
\vspace{-0.2cm}
No evidence of new physics has been found in the 
datasets of $t\bar{t}$ events collected with the CDF detector
with integrated luminosities up to 1-1.7 $fb^{-1}$.
A rich variety of analysis methodologies are in placed
for the increased incoming data. The Tevatron Run 2 aims 
for 6-8 $fb^{-1}$ by the end of 2009.

\vspace{-0.4cm}
\section*{Acknowledgments}
\vspace{-0.2cm}
Thanks to the Top Physics group of the CDF experiment for helpful discussions.
The work of S.C has been supported by a MEC Ram\'{o}n y Cajal contract and a MEC project FPA-2005-25357-E.

%

\end{document}